# 3D Hopping conduction in SnO$_2$ nanobelts

E R Viana*, J C González, G M Ribeiro and A G de Oliveira

Departamento de Física, Universidade Federal de Minas Gerais
Av. Pres. Antônio Carlos, 1500, Pampulha, 31270-901, Belo Horizonte/MG – Brazil.

Email: emilson.fisica@gmail.com; emilsonfisica@ufmg.br;

**Abstract**

An individual tin oxide (SnO$_2$) nanobelt was connected in a two-probe configuration and the electrical transport were measured in darkness from 400 to 5K. We found four intrinsic electrical transport mechanisms through the nanobelt. It starts with Thermal-Activation Conduction (TAC) between 400 and 314K, Nearest-Neighbor Hopping (NNH) conduction between 268 and 115K, and Variable Range Hopping (VRH) conduction below 58K, with a crossover from the 3D-Mott-VRH to the 3D-Efros-Shklovskii-VRH at 16K. We claim that this sequence reveal the three-dimensional nature of the electrical transport in the SnO$_2$ nanobelts, even they are expected to be a one-dimensional system.

**Keywords:** tin oxide, nanobelt, electrical transport, variable range hopping conduction, crossover Mott to Efros-Shklovskii VRH.

___

## 1. Introduction

Metal-oxides semiconductor nanostructures such as tin oxide (SnO$_2$) nanowires and nanobelts have large potential for applications in gas[1] and ultraviolet light[2] sensors. The fundamental aspects of the electrical conduction mechanisms have already been studied for SnO$_2$ in bulk[3], nanocrystals[4], and thin films[5,6]. For nanowires nanobelts the results still remain unclear and controversial.

Small diameter SnO$_2$ nanowires, expected to behave as a one-dimensional (1D) system, show some evidences of three-dimensional (3D) transport. Indeed, a temperature (T) dependence of the electrical resistivity ($\rho$) in the form $\ln\rho \sim T^{-1/2}$ has been observed[7] at low temperatures and interpreted as a 3D-Efros-Shklovskii variable-range hopping[8] (ES-VRH) conduction, instead of an 1D-Mott variable-range[9] (Mott-VRH) conduction. The appearance of a "gap" in the differential conductance curves have been as associated to a Coulomb gap and as the confirmation of the 3D-ES-VRH in SnO$_2$[7] and also in ZnO[10] nanowires. However, the values of the "gap" are too large when compared to the theoretical Coulomb gap of SnO$_2$, and the contact resistance can be also miss-interpreted as a "gap" in the differential conductance curves. The direct transition from the Thermally Activated Conduction (TAC) to the 3D-ES-VRH conduction is also controversial, since the Mott-VRH regime was expected. Similar behavior has also been observed in SnO$_2$ nanocrystalline thin films[5]. Furthermore, Sn$_3$O$_4$ nanobelts also exhibit, in the Mott-VRH regime, 3D hopping conduction at low temperatures [11].

In this letter, we will show that the temperature dependence of the electrical transport in high quality SnO$_2$ nanobelts follows the complete sequence. It starts with the TAC regime and changes sucessively to Nearest-Neighbor Hopping[9] (NNH), Mott-VRH, and finally to ES-VRH, not skipping NNH and Mott-VRH as in ref. [7] and [10]. This sequence actually revels the 3D nature of the electrical transport in the SnO$_2$ nanobelts. In addition, the very good agreement between the experimental and theoretical 3D Mott-ES crossover-temperature confirms the 3D nature of the transport.

## 2. Synthesis and characterization

SnO$_2$ nanobelts were synthesized by the gold-catalyst-assisted VLS method[12], as reported elsewhere[13]. Scanning and transmission electron microscopy (SEM, TEM), presented in Figure 1.(A) and in the inset, respectively, show that high quality prismatic cross-sectioned nanobelts with lateral sizes between 50 to 500 nm can be obtained by this growth method. X-ray diffraction measurements, not shown here, confirms the microscopy results and the tetragonal rutile structure[14] of the nanobelts. The two-probe single nanobelt device, shown in Figure 1.(B), was fabricated by photolithography and standard lift-off process. The electrical resistivity of individual nanobelt was measured, by using a *Oxford®* cryostat CF104 and a *Keithley®* 237 source-meter, in the 5 to 400 K temperature range. Current-voltage I(V) curves confirmed the ohmicity of the electrical contacts.

## 3. Results

Figure 2 shows the resistivity of a nanobelt as a function of temperature. Four conduction mechanisms are clearly observed in the results. At high temperatures (400 to 314K) the resistivity of the nanobelts is dominated by the thermal activation of electrons from shallow donors to the conduction band. The activation energy was calculated and the value of $\Delta E_{TA} = (61.8\pm0.1)$ meV is consistent with data reported for oxygen vacancies in SnO$_2$ [15,16].

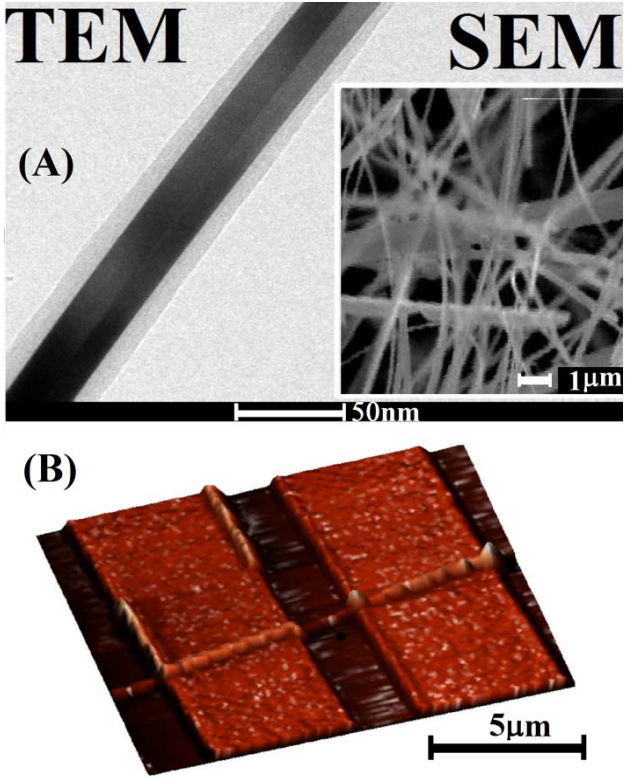

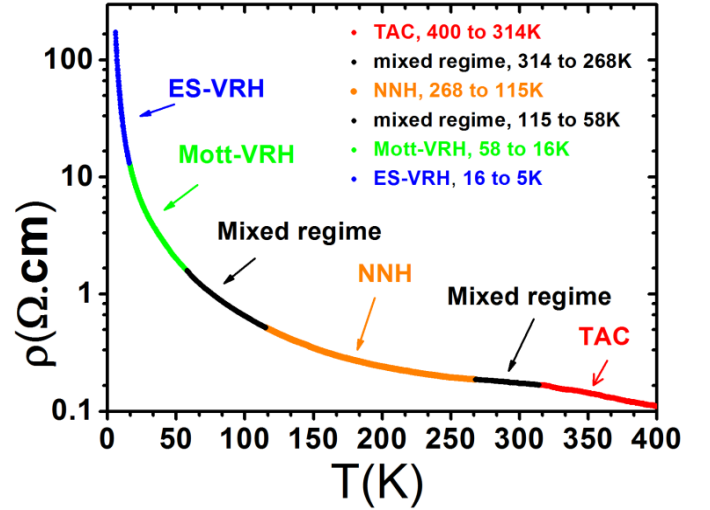

Figure 1. (A) A TEM image of a $SnO_2$ nanobelt covered by a thin amorphous layer. In the inset, the SEM image of the nanobelts. (B) The AFM image of the real two-probe device.

As the temperature decreases the electrical conduction is dominated by hopping mechanisms between donor's sites. At temperatures between 268 and 115K the hopping conduction becomes most probable between nearest ionized impurities neighbors. Electrons can hop between these levels by emitting or absorbing phonons. This NNH[9] resistivity has an expression similar to the TAC resistivity:

$$\rho(T) = \rho_{NNH}. exp\,(\Delta E_{NNH}/k_B T) \qquad (eq.1)$$

where $\Delta E_{NNH}$ corresponds to the magnitude of the potential fluctuation of the donor's levels. The fitting of the data presented in Figure 3 gives $\Delta E_{NNH} = (18.1\pm0.1)$ meV, and $\rho_{NNH} = (85\pm1)\times10^{-3}\Omega.cm$. This activation energy is consistently lower than $\Delta E_{TAC}$, and in agreement with the values reported for $SnO_2$ thin films[6]. As the temperature decreases further, the electrons will hop only to the most probable sites in the neighborhood of the Fermi-level, and the electrical conduction is now dominated by VRH mechanisms. When the Coulomb interactions between electrons is negligible, the Mott-VRH[9] dominates and the resistivity can be expressed as:

$$\rho(T) = \rho_M. exp\,\left(T_M/T\right)^{1/(1+n)} \qquad (eq.2)$$

where $T_M = (\beta_M/\xi^3 N(E_F))$ is Motts's characteristic temperature, $N(E_F)$ is the density of states near the Fermi level, and $\xi$ is the localization length and $\beta_M = 18.1$ [17]. The parameter $n = 1, 2$ or $3$ refers to one, two or three-dimensional (1D, 2D or 3D) hopping conduction.

Figure 2. Electrical resistivity as a function of temperature ramped from 400K to 5K.

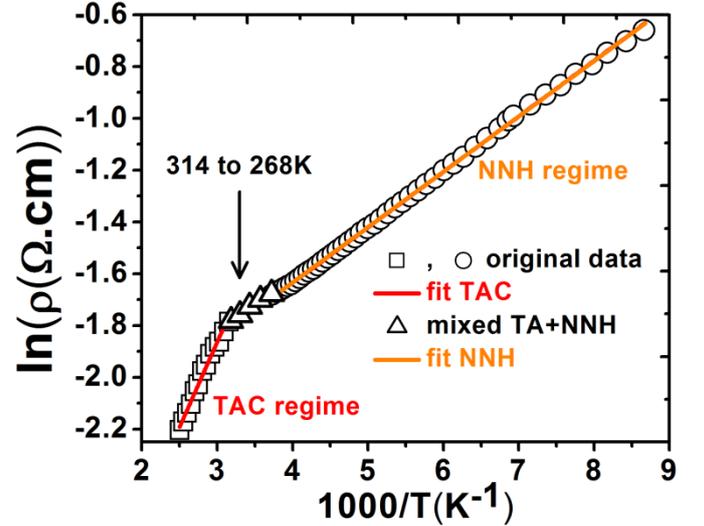

Figure 3. $\ln(\rho)$ vs. $T^{-1}$, showing the TAC (squares) and NNH (circles) regimes.

With the enhancement of the Coulomb interactions between electrons, a soft-gap appears in the density of state (DOS) at the Fermi-level, and the ES-VRH dominates. In this case, the resistivity is given by [8]:

$$\rho(T) = \rho_{ES}. exp\,\left(T_{ES}/T\right)^{3/(3+n)} \qquad (eq.3)$$

Where $(T_{ES} = (\beta_{ES}e^2/\varepsilon k_B \xi)$ is the ES's characteristic temperature, $\varepsilon$ is the dielectric constant, and $\beta_{ES} = 2.8$[8]. Both regimes, the Mott-VRH and the ES-VRH, are presented in Figure 4. The best fit to the experimental data was obtained with $n = 3$, which implies in $T_M = (642\pm3)\times10^2$ K, $\rho_M = (4.8\pm0.1)\times10^{-3}$ Ω.cm, $T_{ES} = (253\pm5)$ K, and $\rho_{ES} = (243\pm1)\times10^{-3}$ Ω.cm.

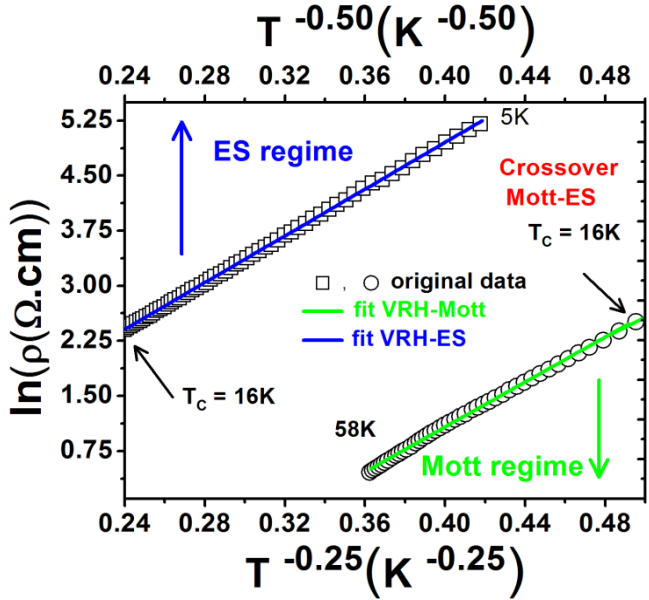

**Figure 4.** The electrical resistivity as a function of temperature in the 3D-Mott-VRH (58 K - 16 K) and 3D-ES-VRH (16 K -5 K) regime.

The Mott-ES transition was observed near 16K, in very good agreement with the theoretical 3D-Mott-ES theoretical crossover temperature that is determined by [18]:

$$T_c = \frac{16 T_{ES}^2}{T_M} = 15.9 \text{ K} \qquad (eq.4)$$

The transition from 3D-Mott-VRH ($\ln\rho \sim T^{-1/4}$) to 3D-ES-VRH ($\ln\rho \sim T^{-1/2}$) observed in the resistivity temperature dependence shows that, in fact, from the electron transport point of view, the $SnO_2$ nanobelts behave as a 3D system, as suggested by the pioneering paper of Yong-Jun Ma et.al.[10], but only confirmed now. This fact is also supported by the very good agreement between theoretical and experimental crossover temperature $T_c$, that also implies in the existence of a soft Coulomb gap of $\Delta_{CG} = k_B.T_c = 1.37$ meV [18]. The presence of the four 3D conduction regimes has been also reported for $SnO_2$ thin films[6] but, the four conduction regimes were however reported with certainly for the first time for $SnO_2$ nanobelts.

## 4. Conclusions

We have studied the temperature dependence of the electrical transport in high quality $SnO_2$ nanobelts synthesized by the Au-assisted VLS method. It was found that the electrical conduction is dominated by a thermal activation process at high temperatures (T > 314K) and it is followed by hopping conduction processes, namely NNH and VRH. Between 314 K and 58 K the electrical transport is dominated by NNH. Between 58 K and 16 K, the temperature dependence of the resistivity follows a $\ln\rho \sim T^{-1/4}$ law, that turns into a $T^{-1/2}$ dependence for temperatures below 16K, characterizing a crossover from the Mott-VRH to ES-VRH, and revealing the 3D nature of the electrical transport in our nanobelt. This 3D nature is also supported by the very good agreement between the experimental and theoretical 3D Mott-ES crossover temperature $T_c$, with which is possible to calculate the soft Coulomb gap of 1.37meV, that are due to the strong Coulomb interaction between hopping carriers at very low temperatures.

## Acknowledgments

The authors' thanks CNPq, CAPES and FAPEMIG, Brazilian official agencies for funding this work.